\documentclass[twocolumn,showpacs,aps,prd,superscriptaddress]{revtex4}

\usepackage{graphicx}
\usepackage{dcolumn}
\usepackage{amsmath}
\usepackage{epsfig}

\RequirePackage{xspace}

\usepackage{relsize}
\def\babar{\mbox{\slshape B\kern-0.1em{\smaller A}\kern-0.1em
    B\kern-0.1em{\smaller A\kern-0.2em R}}}

\def\epem       {\ensuremath{e^+e^-}\xspace}

\def\g     {\ensuremath{\gamma}\xspace}

\def\q     {\ensuremath{q}\xspace}

\def\qqbar {\ensuremath{q\overline q}\xspace}

\def\s     {\ensuremath{s}\xspace}

\def\b     {\ensuremath{b}\xspace}

\def\piz   {\ensuremath{\pi^0}\xspace}

\def\pip   {\ensuremath{\pi^+}\xspace}
\def\pim   {\ensuremath{\pi^-}\xspace}
\def\pipi  {\ensuremath{\pi^+\pi^-}\xspace}

\def\Kbar  {\kern 0.2em\overline{\kern -0.2em K}{}\xspace}

\def\Kz    {\ensuremath{K^0}\xspace}
\def\Kzb   {\ensuremath{\Kbar^0}\xspace}
\def\KzKzb {\ensuremath{\Kz \kern -0.16em \Kzb}\xspace}
\def\Kp    {\ensuremath{K^+}\xspace}
\def\Km    {\ensuremath{K^-}\xspace}

\def\KpKm  {\ensuremath{\Kp \kern -0.16em \Km}\xspace}
\def\KS    {\ensuremath{K^0_{\scriptscriptstyle S}}\xspace} 
 
\def\Kstarz  {\ensuremath{K^{*0}}\xspace}

\def\Kstar   {\ensuremath{K^*}\xspace}

\def\Kstarp  {\ensuremath{K^{*+}}\xspace}

\def\Dbar    {\kern 0.2em\overline{\kern -0.2em D}{}\xspace}

\def\Dz      {\ensuremath{D^0}\xspace}
\def\Dzb     {\ensuremath{\Dbar^0}\xspace}
\def\DzDzb   {\ensuremath{\Dz {\kern -0.16em \Dzb}}\xspace}
\def\Dp      {\ensuremath{D^+}\xspace}
\def\Dm      {\ensuremath{D^-}\xspace}

\def\DpDm    {\ensuremath{\Dp {\kern -0.16em \Dm}}\xspace}

\def\B       {\ensuremath{B}\xspace}
\def\Bbar    {\kern 0.18em\overline{\kern -0.18em B}{}\xspace}
\def\Bb      {\ensuremath{\Bbar}\xspace}
\def\BB      {\ensuremath{B\Bbar}\xspace} 
\def\Bz      {\ensuremath{B^0}\xspace}
\def\Bzb     {\ensuremath{\Bbar^0}\xspace}
\def\BzBzb   {\ensuremath{\Bz {\kern -0.16em \Bzb}}\xspace}
\def\Bu      {\ensuremath{B^+}\xspace}
\def\Bub     {\ensuremath{B^-}\xspace}
\def\Bp      {\ensuremath{\Bu}\xspace}
\def\Bm      {\ensuremath{\Bub}\xspace}

\def\BpBm    {\ensuremath{\Bu {\kern -0.16em \Bub}}\xspace}

\def\BorBbar    {\kern 0.18em\optbar{\kern -0.18em B}{}\xspace}
\def\DorDbar    {\kern 0.18em\optbar{\kern -0.18em D}{}\xspace}
\def\KorKbar    {\kern 0.18em\optbar{\kern -0.18em K}{}\xspace}

\def\chiczero {\ensuremath{\chi_{c0}}\xspace}

\mathchardef\Upsilon="7107
\def\Y#1S{\ensuremath{\Upsilon{(#1S)}}\xspace}

\def\FourS {\Y4S}

\mathchardef\Deltares="7101
\mathchardef\Xi="7104
\mathchardef\Lambda="7103
\mathchardef\Sigma="7106
\mathchardef\Omega="710A

\def\Deltabar{\kern 0.25em\overline{\kern -0.25em \Deltares}{}\xspace}
\def\Lbar{\kern 0.2em\overline{\kern -0.2em\Lambda\kern 0.05em}\kern-0.05em{}\xspace}
\def\Sigbar{\kern 0.2em\overline{\kern -0.2em \Sigma}{}\xspace}
\def\Xibar{\kern 0.2em\overline{\kern -0.2em \Xi}{}\xspace}
\def\Obar{\kern 0.2em\overline{\kern -0.2em \Omega}{}\xspace}
\def\Nbar{\kern 0.2em\overline{\kern -0.2em N}{}\xspace}
\def\Xb{\kern 0.2em\overline{\kern -0.2em X}{}\xspace}

\def\BR         {{\ensuremath{\cal B}\xspace}}

\def\mes        {\mbox{$m_{\rm ES}$}\xspace}

\def\DeltaE     {\mbox{$\Delta E$}\xspace}

\newcommand{\tev}{\ensuremath{\mathrm{\,Te\kern -0.1em V}}\xspace}
\newcommand{\gev}{\ensuremath{\mathrm{\,Ge\kern -0.1em V}}\xspace}
\newcommand{\mev}{\ensuremath{\mathrm{\,Me\kern -0.1em V}}\xspace}
\newcommand{\kev}{\ensuremath{\mathrm{\,ke\kern -0.1em V}}\xspace}
\newcommand{\ev}{\ensuremath{\mathrm{\,e\kern -0.1em V}}\xspace}
\newcommand{\gevc}{\ensuremath{{\mathrm{\,Ge\kern -0.1em V\!/}c}}\xspace}
\newcommand{\mevc}{\ensuremath{{\mathrm{\,Me\kern -0.1em V\!/}c}}\xspace}
\newcommand{\gevcc}{\ensuremath{{\mathrm{\,Ge\kern -0.1em V\!/}c^2}}\xspace}
\newcommand{\mevcc}{\ensuremath{{\mathrm{\,Me\kern -0.1em V\!/}c^2}}\xspace}

\def\invfb   {\ensuremath{\mbox{\,fb}^{-1}}\xspace}

\def\mus  {\ensuremath{\rm \,\mus}\xspace}

\def\mus        {\ensuremath{\,\mu{\rm s}}\xspace}    

\def\to                 {\ensuremath{\rightarrow}\xspace}

\def\pep2{PEP-II}

\newcommand{\chisq}{\ensuremath{\chi^2}\xspace}

\def\gsim{{~\raise.15em\hbox{$>$}\kern-.85em
          \lower.35em\hbox{$\sim$}~}\xspace}
\def\lsim{{~\raise.15em\hbox{$<$}\kern-.85em
          \lower.35em\hbox{$\sim$}~}\xspace}

\def\CP                {\ensuremath{C\!P}\xspace}

\xspace

\newcommand{\figref}[1]{Figure~\ref{fig:#1}}
\newcommand{\tabref}[1]{Table~\ref{tab:#1}}

\def\jetset74   {\mbox{\tt Jetset \hspace{-0.5em}7.\hspace{-0.2em}4}\xspace}

\def\ie                 {{\rm i.e.~}}

\newcommand{\rhop}               {\mbox{$\rho^+$}}

\newcommand{\KstarIp}             {\mbox{$\Kstar(892)^{+}$}}
\newcommand{\BptoKstarIppiz}      {\mbox{$\Bp \to \KstarIp\piz$}}
\newcommand{\KstarIIp}            {\mbox{$\Kstar(1430)^{+}$}}
\newcommand{\fz}                 {\mbox{$f_0$}}
\newcommand{\fI}                 {\mbox{$\fz(980)$}}
\newcommand{\BptofIKp}           {\mbox{$\Bp \to \fI\Kp$}}
\newcommand{\fII}                {\mbox{$f_2(1270)$}}

\newcommand{\fVI}                {\mbox{$f_{\rm X}(1300)$}}
\newcommand{\chicz}              {\mbox{$\chiczero$}}
\newcommand{\BptochiczKp}        {\mbox{$\Bp \to \chicz\Kp$}}
\newcommand{\Kppizpiz}           {\mbox{$\Kp\piz\piz$}}
\newcommand{\BptoKppizpiz}        {\mbox{$\Bp \to \Kp \piz \piz$}}

\def\fscf	{\mbox{$f_{\rm SCF}$}\xspace}

\def\ncand{\ensuremath{31\,673}}
\def\nsig{\ensuremath{1220 \pm 85}}
\def\nsigEffCor{\ensuremath{7427 \pm 518}}

\def\nsigma{\ensuremath{15.6}}
\def\nsigSyg{\ensuremath{10\,\sigma}}
\def\kpipiBFal{\ensuremath{\left(16.2 \pm 1.2 \pm 1.5 \right)\times 10^{-6}}}
\def\kpipiAcp{\ensuremath{-0.06\pm0.06\pm0.04}}

\def\nsigAllCorrKstar{\ensuremath{1078\pm197}}
\def\kstarProdBF{\ensuremath{\left(2.7\pm0.5\pm0.4\right)\times10^{-6}}}
\def\kstarBF{\ensuremath{\left(8.2\pm1.5\pm1.1\right)\times10^{-6}}}
\def\kstarAcp{\ensuremath{-0.06\pm0.24\pm0.04}}

\def\nsigAllCorrfzero{\ensuremath{1186\pm241}}
\def\fzeroProdBF{\ensuremath{\left(2.8\pm0.6\pm0.5\right)\times10^{-6}}}

\def\fzeroAcp{\ensuremath{0.18\pm0.18\pm0.04}}

\def\nsigAllCorrchicz{\ensuremath{245\pm105}}
\def\chiczProdBF{\ensuremath{\left(0.51\pm0.22\pm0.09\right)\times10^{-6}}}
\def\chiczBF{\ensuremath{\left(18\pm8\pm3\pm1\right)\times10^{-5}}}
\def\chiczAcp{\ensuremath{-0.96\pm0.37\pm0.04}}

\newcommand{\onreslumi}  {\mbox{429\invfb}}
\newcommand{\offreslumi} {\mbox{45\invfb}}
\newcommand{\bbpairs}    {\mbox{$470.9\pm2.8$~million}}

\newcommand{\nbb}        {\mbox{$N_{\BB}$}}
\newcommand{\NN}         {\mbox{${\rm NN}_{\rm out}$}}

\newcommand{\splot}    {\mbox{$_s{\cal P}lot$}\xspace}
\newcommand{\sweights} {\mbox{$_s{\cal W}eights$}\xspace}
\newcommand{\sweight}  {\mbox{$_s{\cal W}eight$}\xspace}

\begin{document}



\title{
  {\large 
    \bf Observation of the rare decay {\boldmath $\Bp\to\Kp\piz\piz$} and
    measurement of the quasi-two-body contributions {\boldmath
    $\BptoKstarIppiz$}, {\boldmath $\BptofIKp$} and {\boldmath $\BptochiczKp$}}
}

%
\author{J.~P.~Lees}
\author{V.~Poireau}
\author{V.~Tisserand}
\affiliation{Laboratoire d'Annecy-le-Vieux de Physique des Particules (LAPP), Universit\'e de Savoie, CNRS/IN2P3,  F-74941 Annecy-Le-Vieux, France}
\author{J.~Garra~Tico}
\author{E.~Grauges}
\affiliation{Universitat de Barcelona, Facultat de Fisica, Departament ECM, E-08028 Barcelona, Spain }
\author{M.~Martinelli$^{ab}$}
\author{D.~A.~Milanes$^{a}$}
\author{A.~Palano$^{ab}$ }
\author{M.~Pappagallo$^{ab}$ }
\affiliation{INFN Sezione di Bari$^{a}$; Dipartimento di Fisica, Universit\`a di Bari$^{b}$, I-70126 Bari, Italy }
\author{G.~Eigen}
\author{B.~Stugu}
\affiliation{University of Bergen, Institute of Physics, N-5007 Bergen, Norway }
% ...
\collaboration{The \babar\ Collaboration}
\noaffiliation

\date{\today}

\begin{abstract}
We report an analysis of charmless
hadronic decays of charged \B\ mesons to the final state $\Kp\piz\piz$,
using a data sample of \bbpairs\ \BB\ events collected with the
\babar\ detector at the \FourS\ resonance.
We observe an excess of signal events, with a significance above
10 standard deviations including systematic uncertainties, and measure
the branching fraction and \CP\ asymmetry to be 
${\cal B}\left(\Bp\to\Kp\piz\piz\right) = \kpipiBFal$ and
$A_{\CP}\left(\BptoKppizpiz\right) = \kpipiAcp$,
where the uncertainties are statistical and systematic, respectively.
Additionally, we study the contributions of the 
\BptoKstarIppiz, \BptofIKp, and \BptochiczKp\ quasi-two-body decays.
We report the world's best measurements of the branching fraction and \CP\
asymmetry of the \BptoKppizpiz\ and \BptoKstarIppiz\ channels.
\end{abstract}

\pacs{13.25.Hw, 11.30.Er, 12.39.-x}

\maketitle


\section{Introduction}

Recent measurements of rates and asymmetries in $B\to K\pi$ decays have
generated considerable interest because of possible hints of new physics
contributions~\cite{Aubert:2007hh,:2008zza}. 
Unfortunately, hadronic uncertainties prevent a clear interpretation of these
results in terms of physics beyond the Standard Model (SM).
A data-driven approach involving measurements of all observables in the $B\to
K\pi$ system can in principle resolve the theoretical situation, but much more
precise measurements are
needed~\cite{Fleischer:2008wb,Gronau:2008gu,Ciuchini:2008eh}.

The ratios of tree-to-penguin amplitudes in the related pseudoscalar-vector
decays $B\to \Kstar\pi$ and $B\to K\rho$ are predicted to be two to three
times larger than those in $B\to K\pi$.
Hence, these decays could have considerably larger \CP\ asymmetries and thus
provide useful additional
information~\cite{Chang:2008tf,Chiang:2009hd,Gronau:2010dd}.
In \tabref{status} we review the existing experimental measurements of the
channels in the $B\to \Kstar\pi$ system. 
Improved measurements of the $\Kstarp\piz$~\cite{cc} decay
can be obtained using the full \FourS\ \babar\ dataset.

\begin{table}[htb]
\center
\caption{
  Experimental measurements of $B\to \Kstar\pi$ decays.
  Average values come from HFAG~\cite{Asner:2010qj}.
}
\label{tab:status}
\begin{tabular}{c@{\hspace{5mm}}c@{\hspace{5mm}}c@{\hspace{5mm}}c}
\hline
Mode & ${\cal B}\times10^{6}$ & $A_{\CP}$ & References \\
\hline
$\Kstarp\pim$ & $10.3 \pm 1.1$ & $-0.23 \pm 0.08$ & \cite{Aubert:2007bs,Aubert:2008zu,Garmash:2006fh,:2008wwa} \\
$\Kstarp\piz$ & $6.9 \pm 2.3$ & $0.04 \pm 0.29 \pm 0.05$ & \cite{Aubert:2005cp} \\
$\Kstarz\pip$ & $9.9\,^{+0.8}_{-0.9}$ & $-0.020\,^{+0.067}_{-0.061}$ &
\cite{Aubert:2008bj,Garmash:2005rv} \\
$\Kstarz\piz$ & $2.4 \pm 0.7$ & $-0.15 \pm 0.12 \pm 0.02$ & \cite{Chang:2004um,Aubert:2008zu} \\
\hline
\end{tabular}
\end{table}

The four $\Kstar\pi$ decays populate six $K\pi\pi$ Dalitz plots (the four
$K\rho$ decays also produce four of the same six final states).
To date, Dalitz plot analyses have been performed in the channels
$\Kp\pipi$~\cite{Aubert:2008bj,Garmash:2005rv},
$\KS\pipi$~\cite{Aubert:2009me,:2008wwa} and
$\Kp\pim\piz$~\cite{Aubert:2008zu,Wagner:2010zzb}. 
The first two of these have shown the presence of a poorly-understood
structure, dubbed the $f_{\rm X}(1300)$, in the $\pipi$ invariant mass
distribution. 
A study of the invariant mass spectrum in $\Bp\to\Kp\piz\piz$ decays could help
elucidate the nature of this peak, since even-spin states will populate
both $K\pipi$ and $K\piz\piz$ (assuming isospin symmetry), while odd-spin
states cannot decay to $\piz\piz$.

Knowledge of the dominant contributions to the $\Kp\piz\piz$ Dalitz plot
may also help to clarify the interpretation of the inclusive time-dependent
analyses~\cite{Gershon:2004tk} of $\Bz\to\KS\piz\piz$~\cite{Aubert:2007ub}.
For such $\b\to\s$ penguin-dominated decays the na\"ive Standard Model
expectation is that the time-dependent \CP\ violation parameter should be
given by $S_{\CP} \approx -\eta_{\CP} \sin(2\beta)$, where $\eta_{\CP}$ is
the \CP\ eigenvalue of the final state ($+1$ for $\KS\piz\piz$) and $\beta$
is an angle of the
Cabibbo-Kobayashi-Maskawa~\cite{Cabibbo:1963yz,Kobayashi:1973fv}
unitarity triangle.
Currently, the results for $\Bz\to\KS\piz\piz$ show the largest deviation,
among hadronic $b \to s$ penguin-dominated decays~\cite{Asner:2010qj}, from
the angle $\beta$ measured in charmed decays, albeit with a large
uncertainty.
Such deviations could be caused by new physics, but in order to rule out the
possibility of sizeable corrections to the Standard Model prediction,
better understanding of the population of the $K\piz\piz$ Dalitz plots is
necessary.

In this article, we present the results of a search for the three-body decay
$\Bp\to\Kp\piz\piz$, including short-lived intermediate two-body modes that
decay to this final state.
A full amplitude analysis of the three-body decay would require detailed understanding of effects related to the misreconstruction of signal events, such as the smearing of their Dalitz plot positions.
These effects are significant in the final state under study, which involves two neutral pions.
Therefore, in order to avoid heavy reliance on Monte Carlo (MC) simulations, we do not perform a Dalitz plot analysis, but instead extract information on intermediate modes including narrow resonances ($\Kstarp(892)\piz$, $\fI\Kp$ and $\chiczero\Kp$) by studying the two-body invariant mass distributions.

There is no existing previous measurement of the three-body branching
fraction, but several quasi-two-body modes that can decay to this final state
have been seen, with varying significances.
These include $\Bp\to\fI\Kp$, observed in the
$\fI\to\pipi$ channel~\cite{Aubert:2008bj,Garmash:2005rv} and also seen in
$\fI\to\Kp\Km$~\cite{Aubert:2006nu};
$\Bp\to\fII\Kp$, seen in $\fII\to\pipi$~\cite{Aubert:2008bj,Garmash:2005rv};
and $\Bp\to\Kstarp(892)\piz$, seen in
$\Kstarp(892)\to\Kp\piz$~\cite{Aubert:2005cp}.
The decay $\Bp\to\chiczero\Kp$ has also been
observed with $\chiczero\to\pipi$~\cite{Aubert:2008bj,Garmash:2005rv}
and $\chiczero\to\Kp\Km$~\cite{Aubert:2006nu,Garmash:2004wa}.

\section{Event reconstruction and selection}

The data used in the analysis were collected with the
\babar\ detector~\cite{Aubert:2001tu} at the \pep2\ asymmetric-energy 
\epem\ collider at the SLAC National Accelerator Laboratory. The sample
consists of an integrated luminosity of \onreslumi\ recorded at the \FourS\
resonance (``on-peak'') and \offreslumi\ collected 40\,\mev\ below the
resonance (``off-peak'').  The on-peak data sample contains the full
\babar\ \FourS\ dataset, consisting of \bbpairs\ \BB\ events.

\begin{figure}[!htb]
 \includegraphics[width=.4942\textwidth]{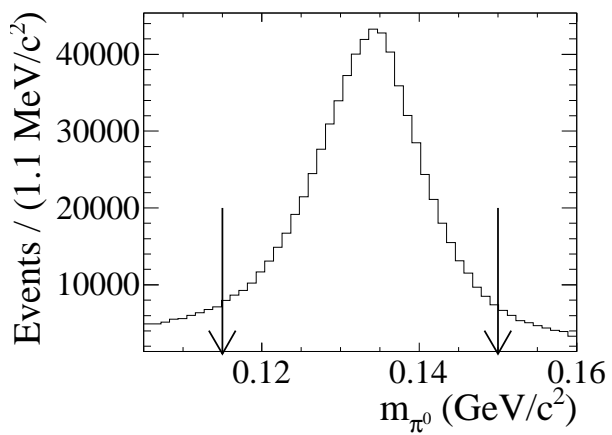} 
 \caption{
   Masses of \piz\ candidates in on-peak data.
   The arrows indicate the selection requirements. 
 }
 \label{fig:piz-mass}
 \end{figure}

We reconstruct $\Bp\to\Kp\piz\piz$ decay candidates by 
combining a \Kp\ candidate with two neutral pion candidates.
The \Kp\ candidate is a charged track with 
transverse momentum above 0.05\,\gevc\ that is consistent with having originated
at the interaction region.  Separation of charged kaons from charged pions
is accomplished with energy-loss information from the tracking subdetectors
and with the Cherenkov angle and number of photons measured by a
ring-imaging Cherenkov detector.
The efficiency for kaon selection is approximately 80\,\% including
geometrical acceptance, while the probability of misidentification of pions
as kaons is below 5\,\% up to a laboratory momentum of 4\,\gevc. 
Neutral pion candidates are formed from pairs of neutral clusters with laboratory
energies above $0.05 \gev$ and lateral moments~\cite{Drescher:1984rt} between
$0.01$ and $0.6$.  We require the mass of the reconstructed \piz\ to be 
within the range $0.115 \gevcc < m_{\g\g} <0.150 \gevcc$ and the
absolute value of the cosine of the decay angle in the \piz\ rest frame to
be less than $0.9$.
\figref{piz-mass} shows the distribution of the mass of neutral pion candidates
in on-peak data. 
Following this selection, when forming the \B\ candidate, the
\piz\ candidates have their masses constrained to the world average
value~\cite{Nakamura:2010zzi}.

We exclude candidates consistent with the $\Bp\to\KS\Kp$,
$\KS\to\piz\piz$ decay chain by rejecting events with a pair of
\piz\ mesons that satisfies $0.40 \gevcc < m_{\piz\piz} < 0.55 \gevcc$.
This veto has a signal efficiency of at least $96\,\%$ for any charmless
resonant decay and of almost $100\,\%$ for nonresonant
$\Bp\to\Kp\piz\piz$ and $\Bp\to\chiczero\Kp$ decays.

Because of the presence of two neutral pions in the final state, there is a
significant probability for signal events to be misreconstructed, 
due to low momentum photons that are replaced by photons from the decay of the
other \B\ meson in the event.  We refer to these as ``self-cross-feed''
(SCF) events, as opposed to correctly reconstructed (CR) events.  Using a
classification based on Monte Carlo information, we find that in
simulated events the SCF fraction depends strongly on the resonant
substructure of the signal, and ranges from $2\,\%$ for
$\Bp\to\chiczero\Kp$ decays to $30\,\%$ for $\Bp\to\fII\Kp$ decays.

In order to suppress the contribution arising from the dominant background,
due to continuum $\epem\to\qqbar\ (\q=u,d,s,c)$ events, we employ a neural
network that combines four variables commonly used to discriminate jet-like
\qqbar\ events from the more spherical \BB\ events. 
The first of these is the ratio of the second-to-zeroth order momentum-weighted
Legendre polynomial moments,
\begin{equation}
\frac{L_2}{L_0} =
\frac
{\sum\limits_{i\in{\rm ROE}}\frac{1}{2}\left(3\cos^{2}{\theta_{i}}-1\right)p_{i}}
{\sum\limits_{i\in{\rm ROE}}p_{i}},
\label{eq:Legendre}
\end{equation}
where the summations are over all tracks and neutral clusters in the event
excluding those that form the \B\ candidate (the ``rest of the event'' or ROE),
$p_{i}$ is the particle momentum and $\theta_{i}$ is the angle between the
particle and the thrust axis of the \B\ candidate.
The three other variables entering the neural network are the absolute value of
the cosine of the angle between the \B\ direction and the beam axis, the
absolute value of the cosine of the angle between the \B\ thrust axis and the
beam axis, and the absolute value of the output of a neural network used for
``flavor tagging'', \ie, for distinguishing \B\ from \Bb\ decays using
inclusive properties of the decay of the other \B\ meson in the
$\FourS\to\B\Bbar$ event~\cite{Aubert:2009yr}.
The first three quantities are calculated in the center-of-mass (CM) frame.
The neural network is trained on a sample of signal MC and off-peak data. We
apply a loose criterion on the neural network output (\NN), which retains
approximately $90\,\%$ of the signal while rejecting approximately $82\,\%$
of the $\qqbar$ background.

In addition to \NN, we distinguish signal from background events using two
kinematic variables:
\begin{eqnarray}
\mes &=& \sqrt{E_{\rm X}^{2}-{\bf p}^{2}_{\B}} \,, \\
\DeltaE &=& E^{\star}_{\B} - \sqrt{s}/2 \,,
\end{eqnarray}
where
\begin{equation}
E_{\rm X} = \left(s/2 + {\bf p}_i \cdot {\bf p}_{\B}\right)/E_i \,,
\end{equation}
$\sqrt{s}$ is the total CM energy,
$\left(E_{i},{\bf p}_{i}\right)$ and $\left(E_{\B},{\bf p}_{\B}\right)$ are the
four-momenta of the initial \epem\ system and \B\ candidate, respectively, both
measured in the lab frame, while the star indicates the CM frame.
The signal \mes\ distribution for CR events is approximately independent of
the $\Bp\to\Kp\piz\piz$ Dalitz plot distribution and peaks near the \B\ mass
with a resolution of about $3\mevcc$.
We select signal candidates with $5.260 \gevcc < \mes < 5.286 \gevcc$.
The CR signal \DeltaE\ distribution peaks near zero, but has a resolution that
depends on the event-by-event Dalitz plot position, the PDF of which is
{\it a priori} unknown.  Prior to the selection of multiple candidates (see
below), we make the requirement $\left|\DeltaE\right|<0.30\gev$, in order
to retain sidebands for background studies.  However, to avoid possible
biases~\cite{Punzi:2004wh} we do not use \DeltaE\ in the fit described
below and instead apply tighter selection criteria for events entering the
fit, $-0.15 \gev < \DeltaE < 0.05 \gev$.  These criteria have an
efficiency of about $80\,\%$ for signal while retaining only about $30\,\%$ of
the background, both compared to the looser requirement
$\left|\DeltaE\right|<0.30\gev$.

The efficiency for signal events to pass all the selection criteria is
determined as a function of position in the Dalitz plot.
Using an MC simulation in which events uniformly populate
phase space, we obtain an average efficiency of approximately $16\,\%$, though
values as low as $8\,\%$ are found near the corners of the Dalitz plot,
where one of the particles is soft.

An average of $1.3$ \B\ candidates is found per selected event.
In events with multiple candidates we choose the one with the smallest value
of a \chisq\ variable formed from the sum of the \chisq\ values of the two
\piz\ candidate masses, calculated from the difference between the
reconstructed \piz\ mass with respect to the nominal \piz\ mass.
This procedure has been found to select the best reconstructed candidate more
than $90\,\%$ of the time, and does not bias our fit variables.

We study residual background contributions from \BB\ events using MC
simulations. 
We divide these events into four categories based on their shapes in the \mes\
and \DeltaE\ distributions.  The first category comprises two-body modes
(mainly $\Bp\to\Kp\piz$); the second contains three-body modes 
(mainly $\Bp\to\Kstarp(\to\Kp\piz)\gamma$ and $\Bp\to\pip\piz\piz$);
the third and fourth are composed of higher multiplicity decays (many possible
sources with or without intermediate charmed states) with missing particles,
and are distinguished by the absence or presence of a peak in the 
\mes\ distribution, respectively. 
Based on the MC-derived efficiencies, total number of \BB\ events, and known
branching fractions~\cite{Asner:2010qj,Nakamura:2010zzi}, 
we expect $70 \pm 9$, $39 \pm 18$, $1090 \pm 40$ and $170 \pm 30$ events in
the four categories respectively. 

\section{\boldmath Study of the inclusive \BptoKppizpiz\ decay}

To obtain the $\Bp\to\Kp\piz\piz$ signal yield, we perform an
unbinned extended maximum likelihood fit to the candidate events using
two input variables: \mes and \NN.
For each component $j$ (signal, \qqbar\ background, and
the four \BB\ background categories), we define a probability density function
(PDF) 
\begin{equation}
  \label{eq:PDF-exp}
  {\cal P}^i_j \equiv
  {\cal P}_j(\mes^i){\cal P}_j({\NN}^i),
\end{equation}
where the index $i$ runs over the selected events.
The signal component is further separated into CR and SCF parts
\begin{equation}
  \label{eq:PDF-sig}
  \begin{array}{rcl}
    {\cal P}^i_{\rm sig} & \equiv &  
    (1 - \fscf) {\cal P}_{\rm CR}(\mes^i){\cal
      P}_{\rm CR}({\cal \NN}^i) + \\
    & \multicolumn{2}{r}{\fscf {\cal P}_{\rm SCF}(\mes^i) 
       {\cal P}_{\rm SCF}({\cal \NN}^i)\, ,}
  \end{array}
\end{equation}
where \fscf\ is the SCF fraction.
The extended likelihood function is
\begin{equation}
  \label{eq:extML-Eq}
  {\cal L} =
  \prod_{k} e^{-n_k}
  \prod_{i}\left[ \sum_{j}n_j{\cal P}^i_j \right],
\end{equation}
where $n_{j(k)}$ is the yield of the event category $j~(k)$.

For the signal, the \mes\ PDFs for CR and SCF are described by an
asymmetric Gaussian with power-law tails and a third-order Chebyshev
polynomial, respectively.
Both CR and SCF \NN\ PDFs are described by nonparametric PDFs (one-dimensional histograms).
We fix the shape parameters of the signal \mes\ PDFs to the
values obtained from the $\Bp\to\Kp\piz\piz$ phase-space MC sample.
The parameters are corrected to account for possible differences between data
and MC simulations, using correction factors determined with a
high-statistics control sample of
$\Bp\to\Dzb\rhop\to\left(\Kp\pim\piz\right)\left(\pip\piz\right)$ decays. 
For the continuum background, we use an ARGUS function~\cite{Albrecht:1990am}
to parameterize the \mes\ shape. The endpoint of the ARGUS function is fixed to
$5.289\gevcc$ whereas the shape parameter is allowed to float in the fit.
The continuum \NN\ shape is modelled with a 20 bin parametric step function,
\ie, a histogram with non-uniform bin width and variable bin content.
One-dimensional histograms are used as nonparametric PDFs to represent all fit
variables for the four \BB\ background components.
The free parameters of our fit are the yields of signal and continuum
background together with the parameters of the continuum \mes\ and \NN\ PDFs. 
All yields and PDF shapes of the four $\BB$ background categories are
fixed to values based on MC simulations. 

The results of the fit are highly sensitive to the value of \fscf, which
depends strongly on the Dalitz plot distribution of signal events and cannot
be determined directly from the fit.
To circumvent this problem, we adopt an iterative procedure.
We perform a fit with \fscf\ fixed to an initial value.
We then construct the signal Dalitz plot from the signal probabilities for
each candidate event (\sweights) calculated with the
\splot\ technique~\cite{Pivk:2004ty}, and determine the corresponding
average value of \fscf.
We then fit again with \fscf\ fixed to the new value, and repeat until
the obtained values of the total signal yield (CR + SCF) and \fscf\ are
unchanged between iterations.
This method was validated using MC and was found to return values of 
\fscf\ that are accurate to within $3\,\%$ of the nominal SCF fraction.
Convergence is typically obtained within three iterations.

We cross-check our analysis procedure using the high statistics control
sample described above.
We impose selection requirements on the $D$ and $\rho$ candidates' invariant
masses: $1.84 \gevcc <m_{\Kp\pim\piz}< 1.88 \gevcc$ and 
$0.65 \gevcc <m_{\pip\piz}< 0.85 \gevcc$.
We fit the on-peak data with a likelihood function that includes components for the
control sample, all \BB\ backgrounds, and \qqbar.
We find a yield that is consistent with
expectation based on the world-average branching fractions~\cite{Nakamura:2010zzi}. 

\begin{figure*}[!htb]
 \includegraphics[width=.4942\textwidth]{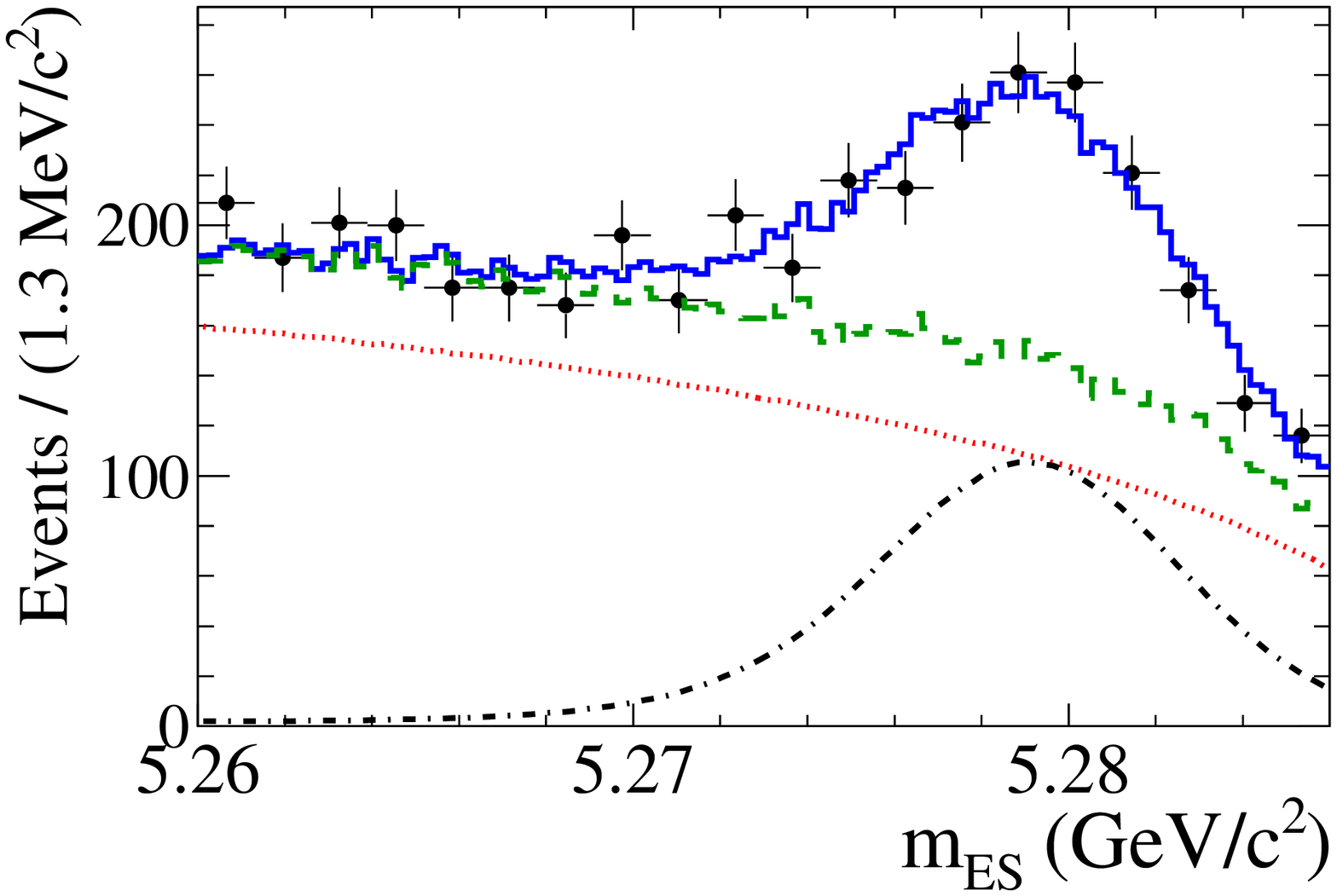} 
 \includegraphics[width=.4942\textwidth]{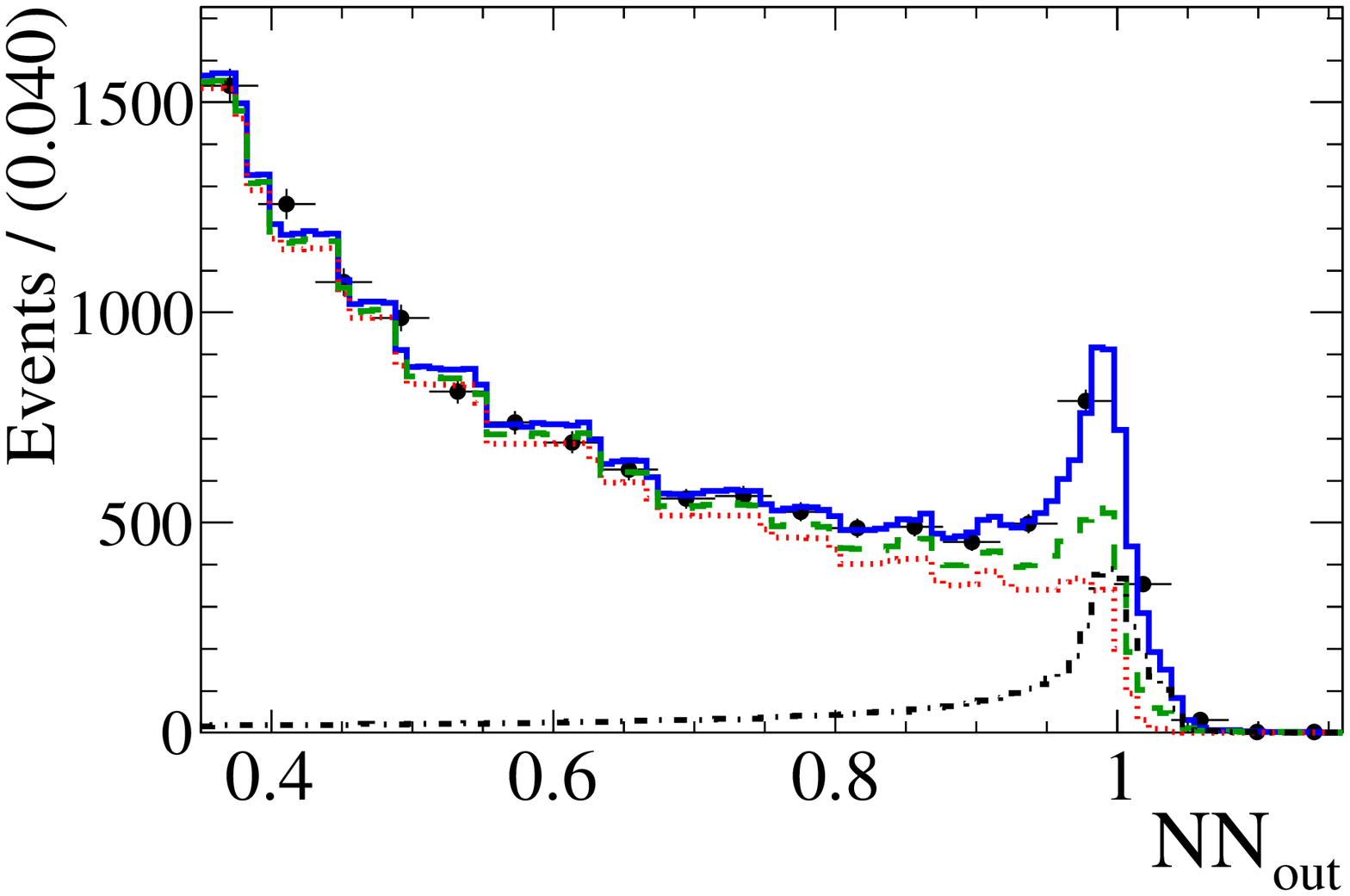}
 \caption{
   Projections of candidate events onto \mes\ (left) and \NN (right),
   following requirements on the other fit variable that enhance signal
   visibility. These requirements retain $60\%$ of signal events for the
   \mes\ plot and $87\%$ of events for the \NN\ plot.
   Points with error bars show the data, the solid (blue) lines the
   total fit result, the dashed (green) lines the total background  
   contribution, and the dotted (red) lines the \qqbar\ component. The
   dash-dotted lines represent the signal contribution.
 }
 \label{fig:signal-project}
 \end{figure*}

We apply the fit method described above to the \ncand\ selected candidate
$\Bp\to\Kp\piz\piz$ events.  Convergence is obtained after four iterations
with a yield of \nsig\ signal events and a SCF fraction of $9.7\,\%$.
The results of the fit are shown in Fig.~\ref{fig:signal-project}.
The statistical significance of the signal yield, 
given by $\sqrt{2\Delta\ln{\cal L}}$ where $\Delta\ln{\cal L}$ is the
difference between the
negative log likelihood obtained assuming zero signal events and that at its
minimum, is \nsigma\ standard deviations ($\sigma$).
Including systematic uncertainties (discussed below), the significance is
above \nsigSyg.

\begin{figure*}[!htb]
   \includegraphics[width=.4942\textwidth]{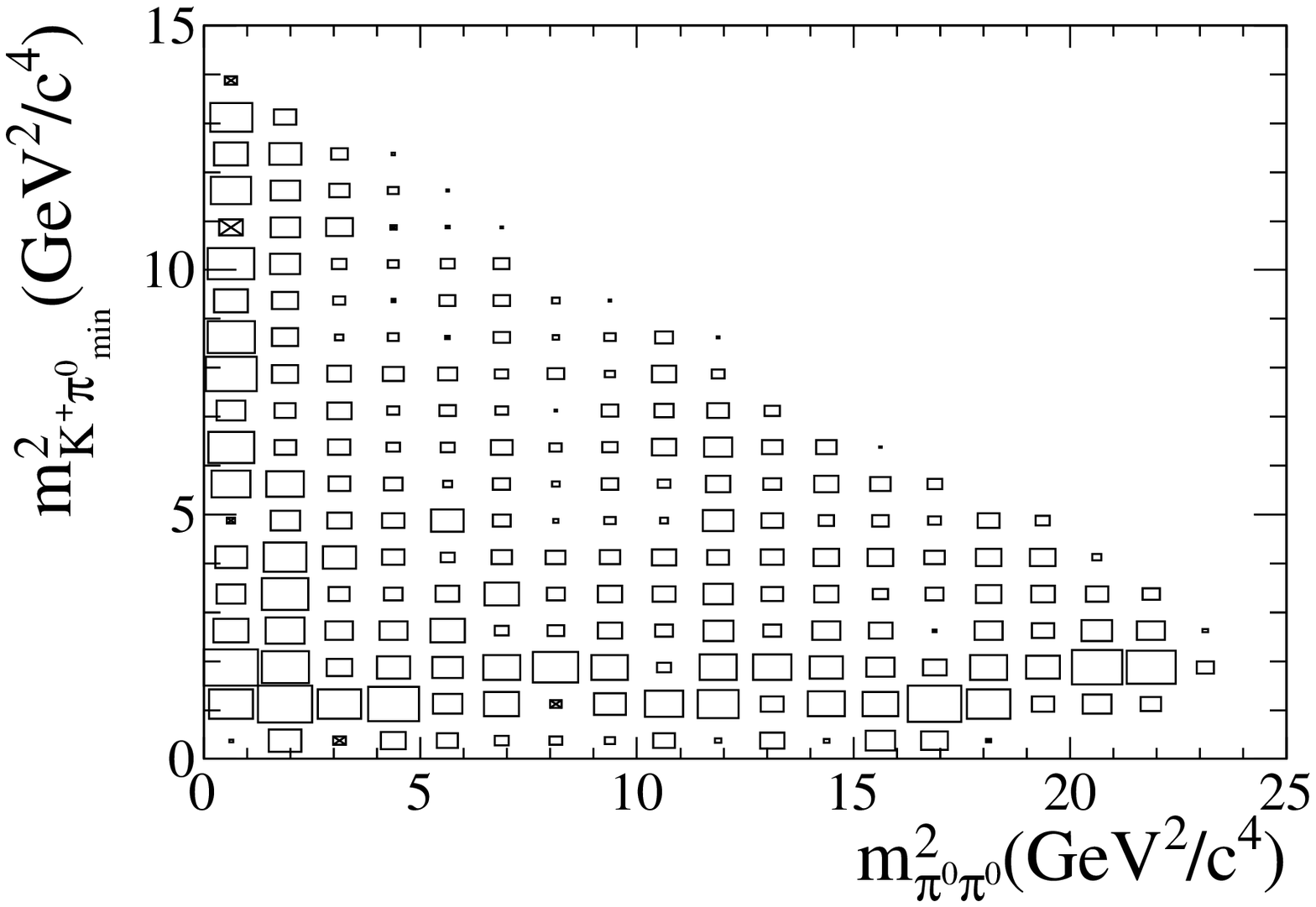} 
   \includegraphics[width=.4942\textwidth]{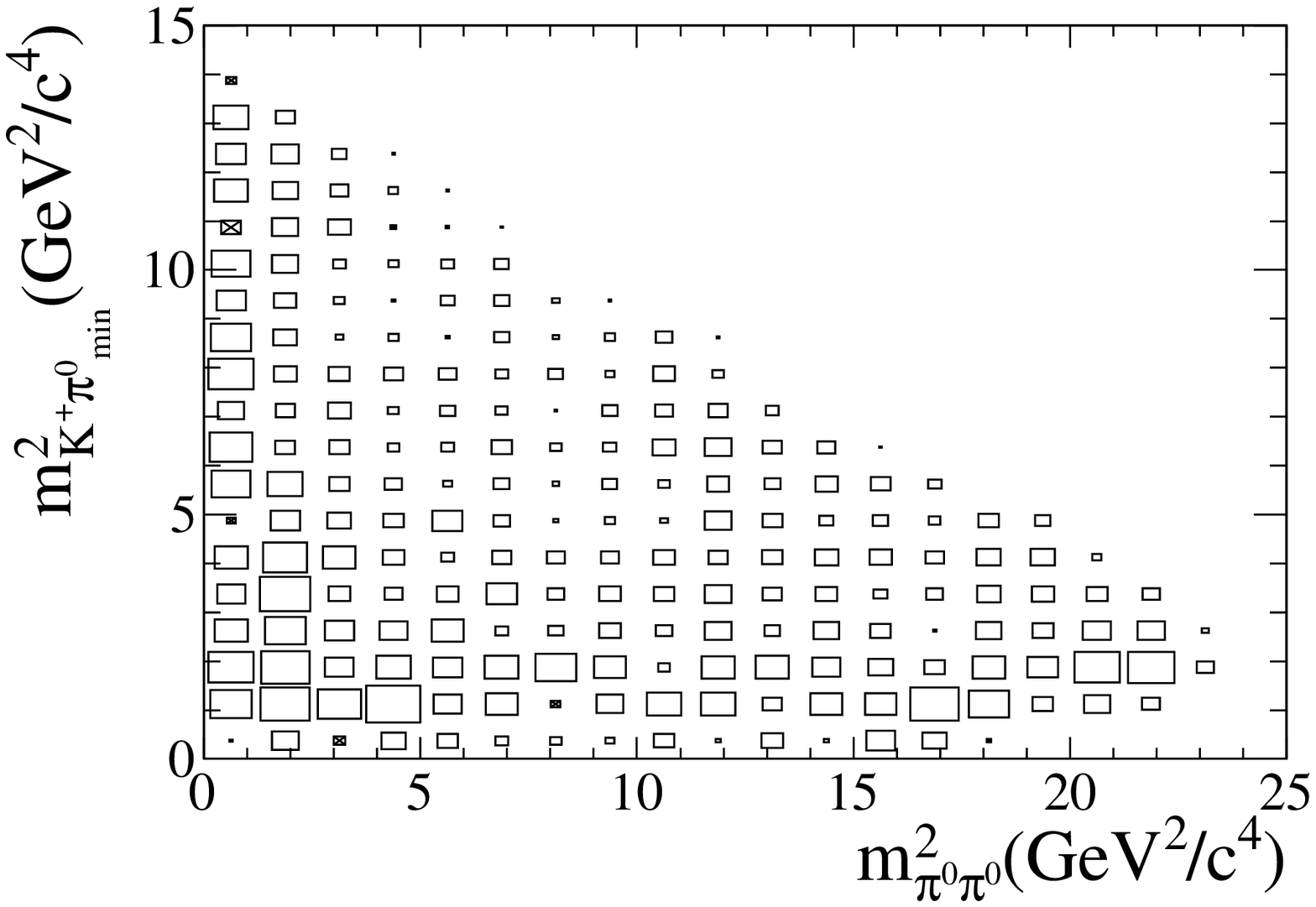}
   \caption{
     Signal Dalitz plot distributions obtained using \sweights\ before
     (left) and after (right) efficiency correction.
     In order to define a unique position for each event, the Dalitz plot
     is shown as $m^2_{\Kp\piz_{\rm min}}$ {\it vs.} $m^2_{\piz\piz}$,
     where $m^2_{\Kp\piz_{\rm min}}$ is the smaller of the two $\Kp\piz$
     invariant masses. 
     Resonance bands are visible for \KstarIp\ at $m^2_{\Kp\piz_{\rm min}} \sim 0.8~{\rm GeV^{2}/c^{4}}$,
     \fI\ at $m^2_{\piz\piz} \sim 1~{\rm GeV^{2}/c^{4}}$ and \chicz\ at $m^2_{\piz\piz} \sim 12~{\rm GeV^{2}/c^{4}}$.}
   \label{fig:signal-DP}
 \end{figure*}

To obtain the $\Bp\to\Kp\piz\piz$ branching fraction using the result of
the fit, we form, for each event, the ratio of the signal
\sweight\ and the efficiency determined from its Dalitz-plot position.
Summing these ratios over all events in the data sample, we obtain an
efficiency-corrected signal yield of \nsigEffCor\ events. 
The \sweight\ calculation accounts for the fixed \BB\ backgrounds~\cite{Pivk:2004ty}.
The Dalitz plot distributions obtained before and after
applying the efficiency correction are shown in Fig.~\ref{fig:signal-DP}.
We apply further corrections for the effect of the \KS\ veto ($98\%$);
differences between data and MC for the \piz\ reconstruction efficiency,
determined from control samples of $\tau$ decays as a function of \piz\
momentum ($95.7\%$);
and a bias in the fitted signal yield (raw bias $44$ events), as determined
from Monte Carlo pseudoexperiments generated with a signal component with
the same values of the yield and SCF fraction as found in the fit to data.
Finally, we divide by the total number of \BB\ events in the data sample
to obtain our measurement of the branching fraction
${\cal B}\left(\Bp\to\Kp\piz\piz\right) = \kpipiBFal$,
where the first uncertainty is statistical and the second is systematic.

The systematic uncertainty includes contributions from the PDF shapes; the
fixed \BB\ background yields; the estimation of the SCF fraction; intrinsic fit
bias; selection requirements; and the number of \BB\ pairs in the data sample.
Here we provide further details on each of these sources of systematic
uncertainty and describe briefly how each is evaluated.
A combined uncertainty for CR signal and \BB\ background \NN\ PDF shapes
(4.9\,\%) is evaluated using uncertainties in the data/MC ratio determined from
the $\Bp\to\Dzb\rhop\to\left(\Kp\pim\piz\right)\left(\pip\piz\right)$ control
sample and applying them simultaneously to the CR signal and \BB\ background
\NN\ PDFs.
The same control sample is used to evaluate the uncertainties in CR signal
\mes\ PDF shapes (0.8\,\%).
The uncertainty in the SCF fraction (2.5\,\%) is estimated by varying the value
used in the fit within a range of uncertainty determined from Monte Carlo
pseudoexperiment tests of our iterative fitting procedure.
Uncertainties in the SCF signal \mes\ and \NN\ PDF shapes (1.7\,\% and
0.7\,\%, respectively) are evaluated by considering a range of SCF shapes
corresponding to different signal Dalitz plot distributions.
An uncertainty in the correction due to fit bias (1.9\,\%) is assigned, which
corresponds to half the correction combined in quadrature with its error.
Uncertainties in the \BB\ background \mes\ PDF shapes due to data/MC
differences (1.6\,\%) are evaluated by smearing the PDFs with a Gaussian with
parameters determined from the $\Bp\to\Dzb\rhop$ control sample.
The uncertainties in the \BB\ background PDFs due to finite MC statistics (0.8\,\%)
are determined by varying the contents of the bins of the histograms used to
describe the PDFs within their errors.
Uncertainties in the fixed \BB\ background yields (1.4\,\%) are evaluated by
varying these yields within their uncertainties.
Contributions to the uncertainty in the selection efficiency arise from
the \DeltaE\ (4.0\,\%) and \NN\ (3.0\,\%) selection requirements,
neutral pion reconstruction (2.8\,\%),
the \KS\ veto correction (2.0\,\%),
kaon identification (1.0\,\%) and
tracking (0.4\,\%);
The uncertainty in the number of \BB\ pairs in the data sample is 0.6\,\%.
Including only systematic uncertainties that affect the fitted yield, 
the total is 6.5\,\%. 
The total systematic uncertainty on the branching fraction is
9.0\,\%. \tabref{systematics} summarizes the systematic contributions.

\begin{table}
  \caption{
    Summary of systematic uncertainties for the inclusive branching fraction
    measurement.
  }
  \label{tab:systematics}
  \begin{tabular}{l|c}
    \hline
    Source & Uncertainty \\
    \hline
    CR signal and \BB\ background \NN\ PDFs & 4.9\,\% \\
    CR signal \mes\ PDF & 0.8\,\% \\
    SCF fraction & 2.5\,\% \\
    SCF signal \mes\ PDF & 1.7\,\% \\
    SCF signal \NN\ PDF & 0.7\,\%\\
    Fit bias & 1.9\,\% \\
    \BB\ background \mes\ PDFs & 1.6\,\% \\
    \BB\ background PDFs (MC statistics) & 0.8\,\% \\
    \BB\ background yields & 1.4\,\% \\
    \hline
    Subtotal & 6.5\,\% \\
    \hline
    \DeltaE\ selection efficiency & 4.0\,\% \\
    \NN\ selection efficiency     & 3.0\,\% \\
    Neutral pion efficiency & 2.8\,\% \\
    \KS\ veto & 2.0\,\% \\
    Particle identification efficiency & 1.0\,\% \\
    Tracking efficiency & 0.4\,\% \\
    \nbb & 0.6\,\% \\
    \hline
    Total & 9.0\,\% \\
    \hline
  \end{tabular}
\end{table}

The \CP\ asymmetry is measured as
\begin{equation}
A_{\CP}=\frac{N_{\Bm}-N_{\Bp}}{N_{\Bm}+N_{\Bp}} \,,
\end{equation}
where $N_{\Bp\left(\Bm\right)}$ is the number of events from \BptoKppizpiz\
(\CP\ conjugate decay) and is obtained by including in the above-described
fit the value of the kaon charge. The fit returns an asymmetry of
$A_{\CP}=\kpipiAcp$.
Most of the systematic uncertainties that affect the branching fraction cancel
in the asymmetry.  However, the following sources are considered and evaluated for the $A_{\CP}$ measurement.
Detector-induced asymmetries have been studied in previous similar
analyses~\cite{Aubert:2005cp,Aubert:2008bj} and found to be small (0.5\,\%).
We evaluate the possibility that our selection induces an asymmetry by measuring
the \CP\ asymmetry in the $\Bp\to\Dzb\rhop$ control sample (3.0\,\%), where
none is expected.
The \BB\ background asymmetries are fixed in our fit; the uncertainty from this is
evaluated (1.8\,\%) by varying these by a weighted average of the
\CP\ asymmetries of the contributing \BB\ decays.
Finally the fit bias is estimated from MC pseudoexperiments (1.2\,\%).

\section{Study of quasi-two-body contributions}

\begin{figure*}[!htb]
  \includegraphics[width=.329\textwidth]{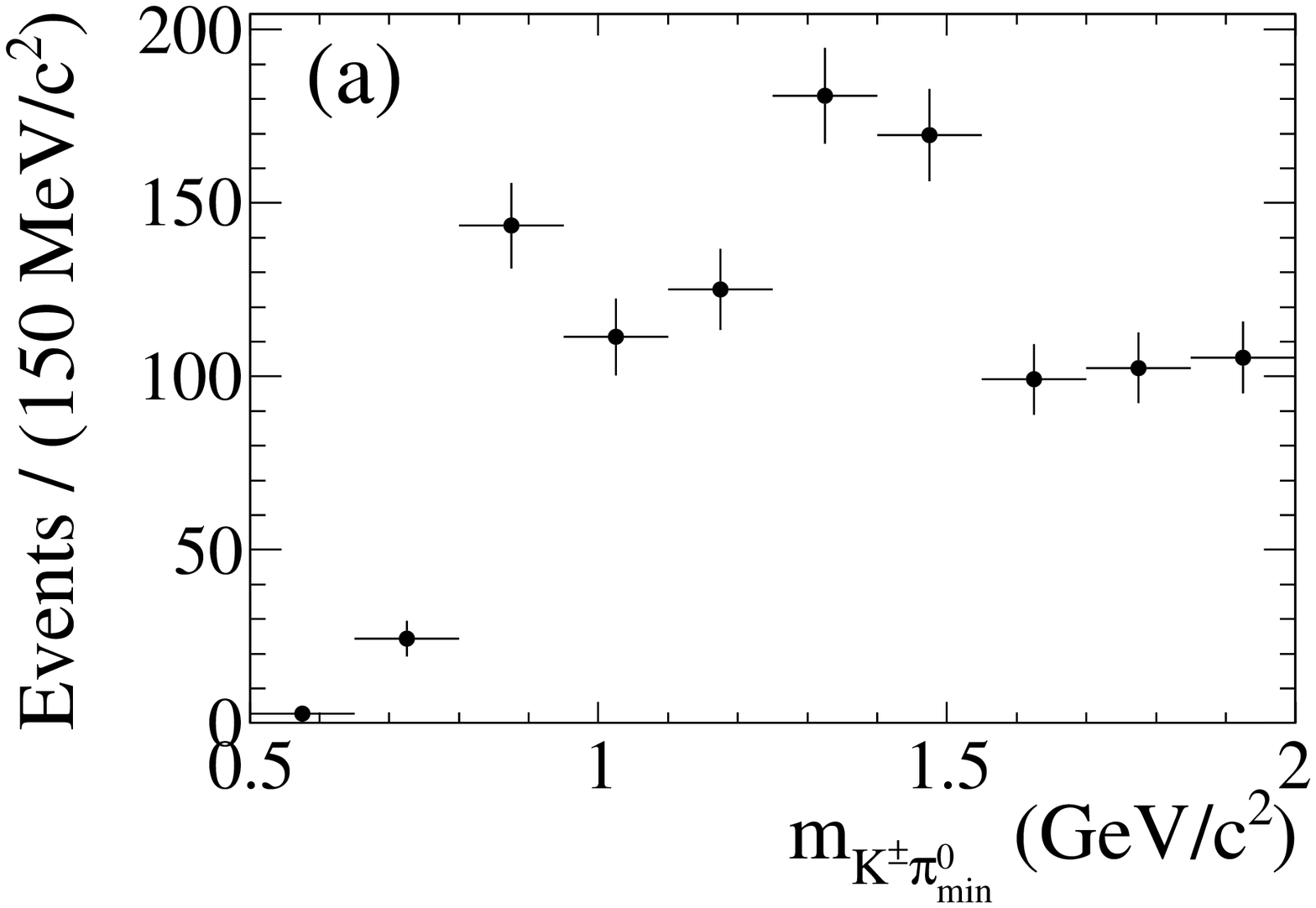}
  \includegraphics[width=.329\textwidth]{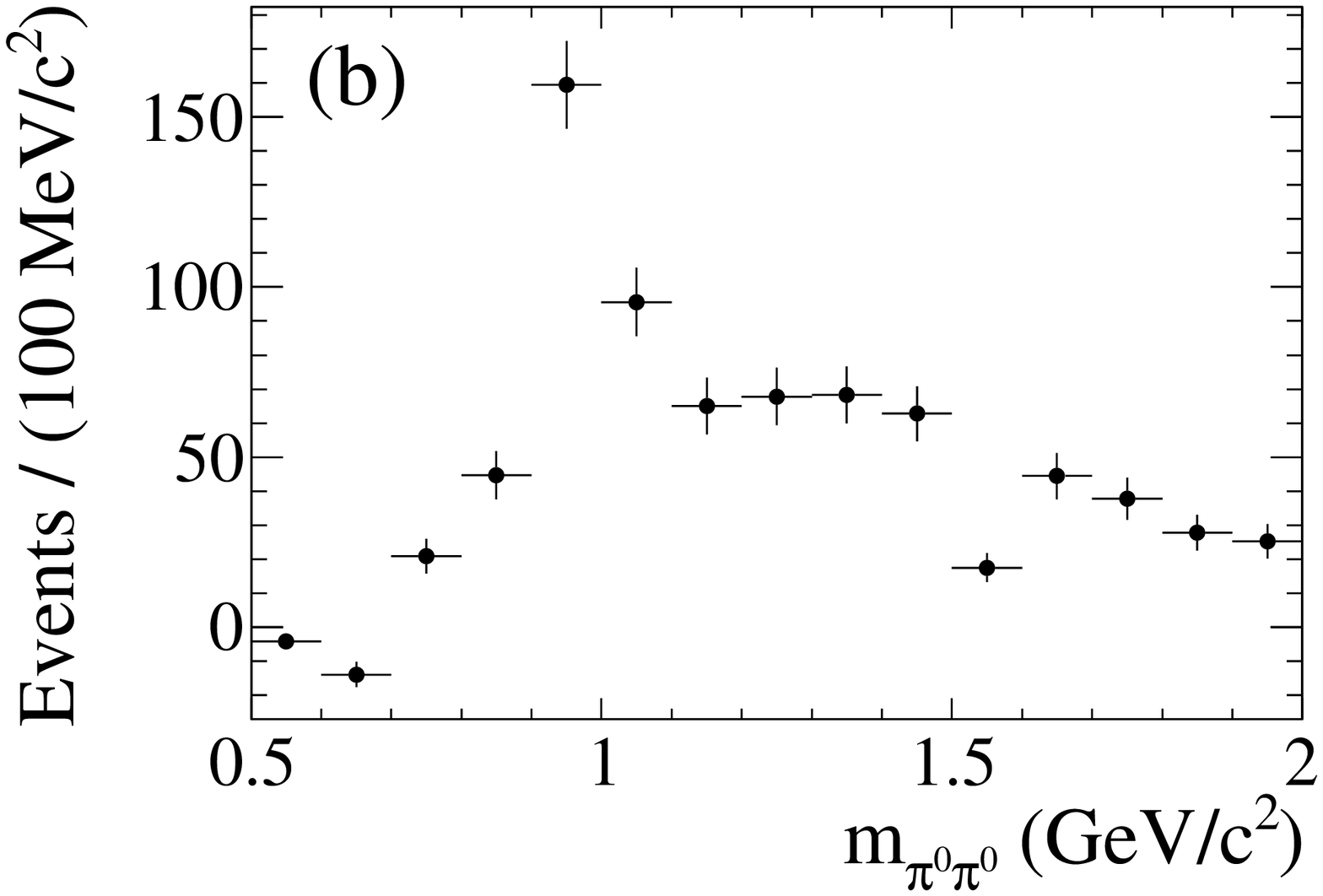} 
  \includegraphics[width=.329\textwidth]{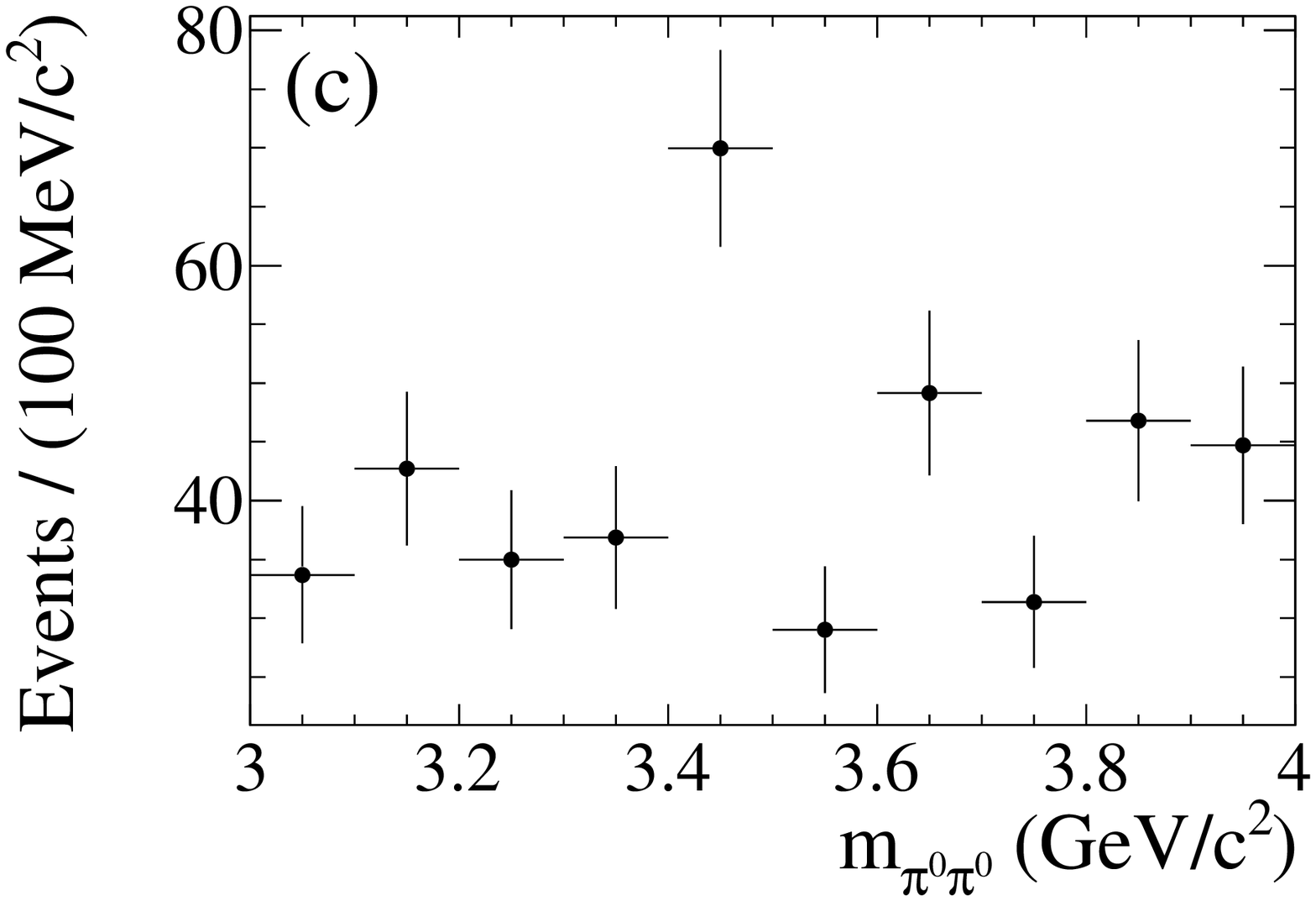}
  \caption{
    Signal \splot\ distributions not corrected for efficiency for 
    (a) $0.5\gevcc<m_{\Kp\piz_{\rm min}}<2.0\gevcc$,
    (b) $0.5\gevcc<m_{\piz\piz}<2.0\gevcc$
    and (c) $3.0\gevcc<m_{\piz\piz}<4.0\gevcc$.
    $m_{{\Kp\piz}_{\rm min}}$ is the $\Kp\piz$ combination with lower
    invariant mass.  Excesses of events in the \fI, \chicz, \KstarIp, and
    \KstarIIp\ mass regions are clearly visible. 
  }
 \label{fig:mass-splot}
 \end{figure*}

We use the \splot\ distributions obtained from the fit and projected onto
the Dalitz plot axes to search for peaks from intermediate resonances.
These projections are shown for both $\Kp\piz$ and $\piz\piz$ invariant masses
in Fig.~\ref{fig:mass-splot}. 
Signal peaks from \KstarIp, \fI\ and \chicz\ are clearly observed.
We do not see any enhancement that could be attributed to the \fVI,
though the $\piz\piz$ invariant mass distribution contains a pronounced dip
around $1550\mevcc$ that could arise from interference between various
resonances in this region.
A broad peak around $1400\mevcc$ in the $\Kp\piz$ invariant mass distribution could be due to contributions from spin-0 and/or spin-2 $\Kstar(1430)^{+}$ states.

\begin{figure*}[!htb]
  \includegraphics[width=.329\textwidth]{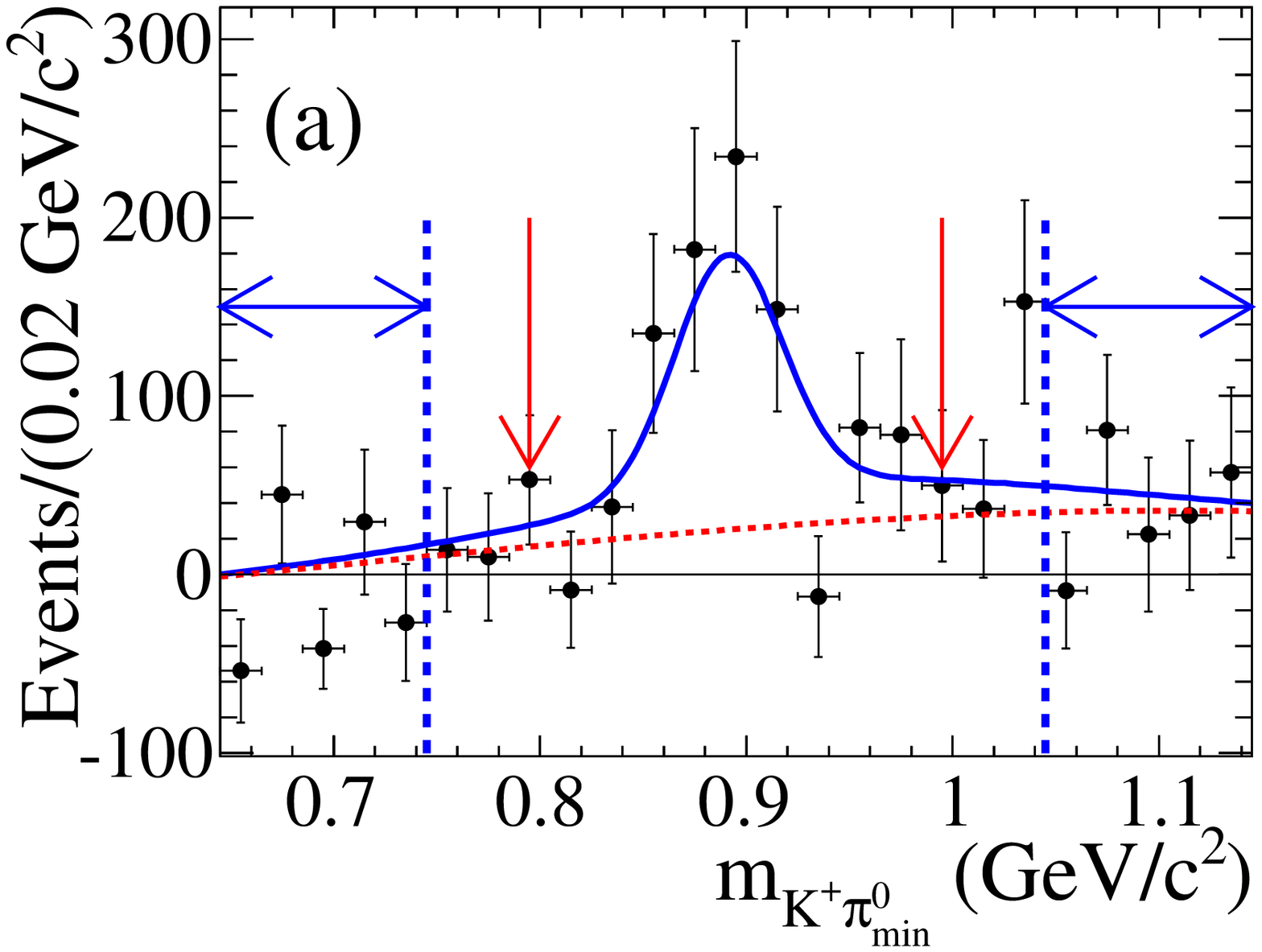} 
  \includegraphics[width=.329\textwidth]{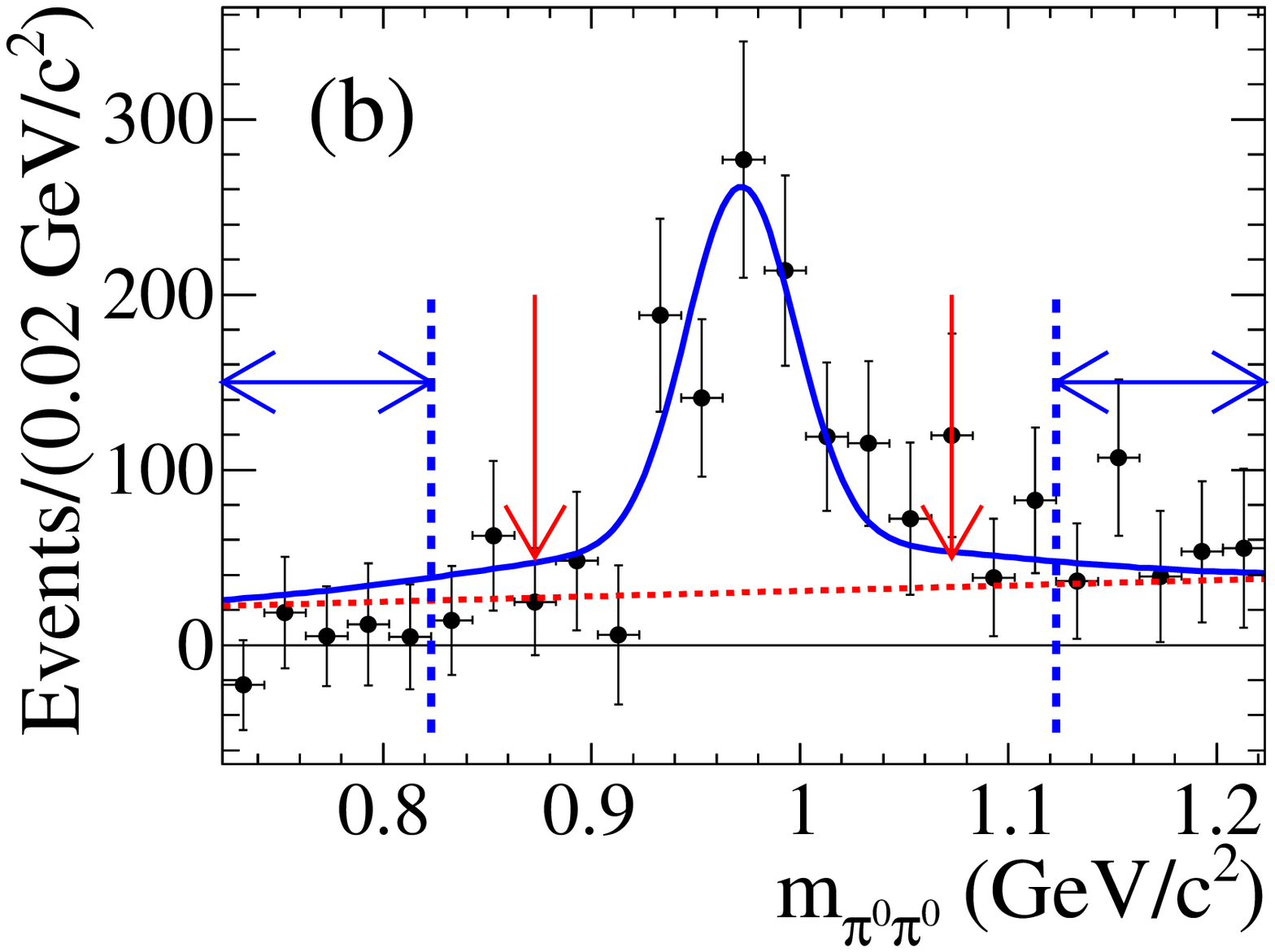}
  \includegraphics[width=.329\textwidth]{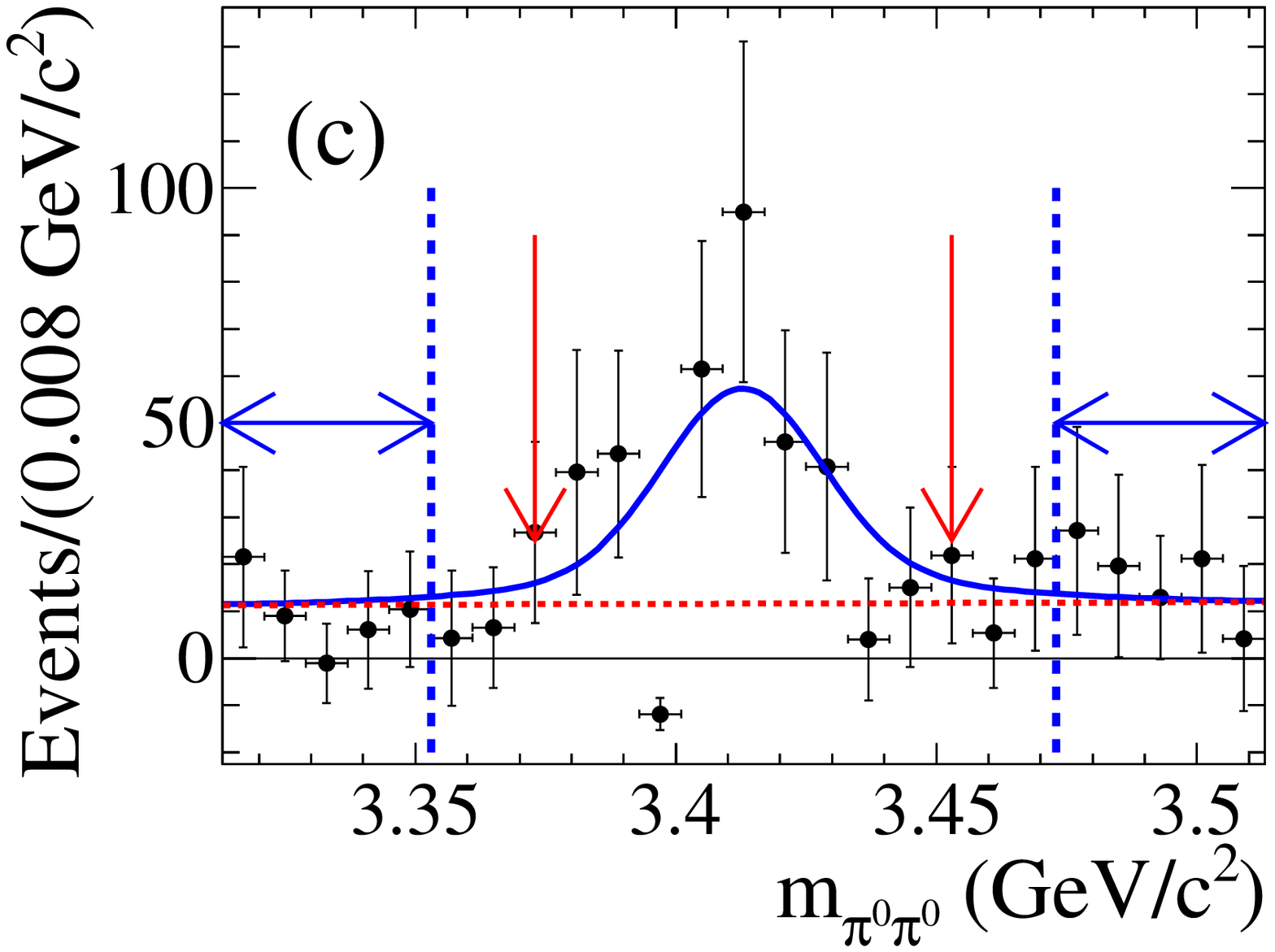}
  \caption{
    Efficiency-corrected signal (vertical red arrows) and sideband
    (horizontal blue arrows) regions around (a) \KstarIp, (b) \fI\
    and (c) \chicz\ invariant mass peaks.  The curves show the results of the
    fit used to cross-check the procedure, for the total (blue continuous
    line) and background-only (dashed red line) components.
  }
  \label{fig:q2b-signal}
\end{figure*}

The numbers of signal events for the quasi-two-body contributions are
determined by defining signal regions around the peaks of the resonances. 
Efficiency-corrected \sweights\ are summed in the same way as used to measure
the inclusive branching fraction.  To estimate contributions from
nonresonant and resonant \BptoKppizpiz\ decays other than the
quasi-two-body decays under consideration (which we refer to as background
in this section), the same procedure is applied to sidebands on either side
of each signal region in the two-particle invariant mass.  The background
yields are estimated as the normalized averages of the two sidebands'
yields and are subtracted from the efficiency-corrected yields in
the signal regions.
The signal and sideband regions are illustrated by arrows for each of the three
quasi-two-body modes in Fig.~\ref{fig:q2b-signal}.
We use this approach rather than a full Dalitz plot analysis since
the latter would require more detailed understanding of the properties of SCF
events. 
Our method does, however, suffer from systematic uncertainties (evaluated below)
due to other contributions to the Dalitz plot and possible interference effects.
This precludes its use for studying quasi-two-body decays via broad resonances.
We have validated our approach using ensembles of MC simulations with
varying mixtures of resonant substructure, and find that in all cases we
are able to correctly obtain the true values of the branching fractions of
the quasi-two-body decays under study, which all have narrow intermediate
states under study.

Fits to the efficiency-corrected invariant mass distributions are used to
cross-check the results of the subtraction method.
In these fits we describe the signal distributions with double-Gaussian
functions, with parameters obtained from MC simulations, and the background
shapes with polynomials.
The two methods yield consistent results, both in MC simulations and in data.

After background subtraction we obtain efficiency-corrected signal yields of 
\nsigAllCorrKstar\ for \BptoKstarIppiz, 
\nsigAllCorrfzero\ for \BptofIKp,
and \nsigAllCorrchicz\ for \BptochiczKp. 
We correct each yield for the inefficiency of the corresponding 
signal region selection, obtained from Monte Carlo simulations.  
Finally the yields are corrected: 
(i) for bias, estimated from Monte Carlo pseudoexperiments;
(ii) for \piz\ efficiency, using the momentum distributions of both \piz\
mesons from a Monte Carlo cocktail reflecting the yields obtained in data;
(iii) in the case of the \KstarIp\ yield only, for the \KS\ veto.
Finally, we divide by the number of \BB\ pairs to obtain the product
branching fractions 
\begin{equation}
\begin{array}{lcr}
  \multicolumn{2}{l}{
    {\cal B}\left(\BptoKstarIppiz\right)\times
    {\cal B}\left(\KstarIp\to\Kp\piz\right) =} & \\
  & \multicolumn{2}{r}{\kstarProdBF\,,} \\
  \multicolumn{2}{l}{
    {\cal B}\left(\BptofIKp\right)\times
    {\cal B}\left(\fI\to\piz\piz\right) = } & \\
  & \multicolumn{2}{r}{\fzeroProdBF\,,} \\
  \multicolumn{2}{l}{
    {\cal B}\left(\BptochiczKp\right)\times
    {\cal B}\left(\chicz\to\piz\piz\right) =} & \\
  & \multicolumn{2}{r}{\chiczProdBF\,,}
\end{array}
\end{equation}
where the first uncertainties are statistical and the second systematic.
The sum of these contributions does not saturate the inclusive branching
fraction, indicating significant contributions from other sources, as is
also clear from Fig.~\ref{fig:signal-DP} and Fig.~\ref{fig:mass-splot}, and expected from the results of studies of $\Bp\to\Kp\pipi$ decays~\cite{Aubert:2008bj,Garmash:2005rv}.

Systematic uncertainties include all the same sources in the same relative
amounts as evaluated for the inclusive decay except for fit bias, \KS\ veto and 
\piz\ efficiency, which are evaluated separately for each quasi-two-body mode.
We also evaluate the following additional contributions.
The uncertainty due to the method of background subtraction (3.5\,\% for
$\KstarIp\piz$, 11.9\,\% for $\fI\Kp$ and 13.5\,\% for $\chicz\Kp$) is obtained
by comparing the nominal results with those from obtained with alternative
sideband regions.
We evaluate the potential effect of interference (10.0\,\%, for $\fI\Kp$ only)
using toy Monte Carlo events generated for a Dalitz plot model containing
\fI\ and nonresonant components with relative magnitudes obtained from the fit
results, and a relative phase sampled in a range that gives distributions
consistent with the data. 
Finally we consider possible data/MC differences affecting the signal region
efficiency correction (5.6\,\% for $\KstarIp\piz$,
3.8\,\% for $\fI\Kp$, and 0.4\,\% for $\chicz\Kp$) determined from the change
in the result when the SCF fraction is varied in Monte Carlo events.
The $\KstarIp\piz$ and $\chicz\Kp$ branching fraction measurements are not
affected by systematics due to interference.
For the former, effects of interference with $\Kp\piz$ S-wave contributions cancel when integrated over the part of the Dalitz plot inside the signal mass window, while P-wave contribution are not expected based on studies of related decays~\cite{Aubert:2008bj,Garmash:2005rv}.
For the latter, the small width implies that interference will be negligible.
A list of the systematic uncertainty contributions is given in
\tabref{q2b-systematics}. 

\begin{table}
  \caption{
    Summary of systematic uncertainties for the branching fraction
    measurement of the quasi-two-body resonances. 
    The breakdown of the systematics affecting the inclusive branching
    fraction measurement is given in \tabref{systematics}.
  }
  \label{tab:q2b-systematics}
  \begin{tabular}{l|ccc}
    \hline
    Source & \multicolumn{3}{c}{Uncertainty (\%)} \\
    & \KstarIp\piz & \fI\Kp & \chicz\Kp	\\
    \hline
    Subtotal from inclusive & 8.1 & 8.1 & 8.1 \\
    \hline
    Background subtraction & 3.5 & 11.9 & 13.5 \\
    Interference & -- & 10.0 & -- \\
    Fit bias & 6.6 & 2.1 & 6.8 \\
    Mass cut efficiency & 5.6 & 3.8 & 0.4 \\
    \piz\ efficiency & 3.1 & 3.5 & 2.6 \\
    \KS\ veto & 2.0 & -- & -- \\
    \hline
    Total & 12.9 & 18.4 & 17.4 \\
    \hline
  \end{tabular}
\end{table}

To obtain the $B$ decay branching fractions, we correct for 
${\cal B}\left(\KstarIp\to\Kp\piz\right)=1/3$ and 
${\cal
B}\left(\chicz\to\piz\piz\right)=\left(8.4\pm0.4\right)\times10^{-3}\times1/3$~\cite{Nakamura:2010zzi},
where the factors of $1/3$ are due to isospin. 
(The branching fraction of $\fI\to\piz\piz$ is unknown, 
hence we cannot correct for it.)
We obtain
\begin{equation}
\begin{array}{rcl}
  {\cal B}\left(\BptoKstarIppiz\right) & = & \kstarBF\,, \\
  {\cal B}\left(\BptochiczKp\right) & = & \chiczBF\,,
\end{array}
\end{equation}
where the first uncertainty is statistical, the second systematic, 
and the third (for \BptochiczKp) is from the subdecay branching fraction.

We obtain the \CP\ asymmetries of the quasi-two-body modes with the same
method used to obtain the quasi-two-body branching fractions, except we
distinguish the yields of the \Bp\ and \Bm\ decays. 
We obtain the following asymmetries:
\begin{equation}
\begin{array}{c}
  A_{\CP}\left(\BptoKstarIppiz\right)=\kstarAcp\,, \\
  A_{\CP}\left(\BptofIKp\right)=\fzeroAcp\,, \\
  A_{\CP}\left(\BptochiczKp\right)=\chiczAcp\,,
\end{array}
\end{equation}
where the first uncertainty is statistical and the second systematic.
The sources of systematic uncertainty are the same as for the inclusive
\CP\ asymmetry measurement. 
The measurements of \CP\ asymmetries for $\BptofIKp$ and $\BptochiczKp$ are
consistent with the world average values based on decays of the intermediate
resonances to $\pipi$~\cite{Asner:2010qj,Nakamura:2010zzi}.
The $\BptochiczKp$ result has a large and non-Gaussian uncertainty and its
difference from zero is not statistically significant.

\section{Conclusion}

In summary, using the full \babar\ data sample of \onreslumi\ collected at the
$\FourS$ resonance, we observe charmless hadronic decays of charged
$\B$ mesons to the final state $\Kp\piz\piz$.
The signal has a significance above \nsigSyg\ after taking systematic effects
into account.  

We study the Dalitz plot distribution of the signal events, and do not see any
excess that could be attributed to the \fVI. However, due to the possible
complicated interference pattern, we cannot draw any strong conclusion
about this state from our analysis.
We measure the product branching fractions and direct \CP\ asymmetry
parameters of the quasi-two-body modes with narrow resonance peaks in the
$\Kp\piz\piz$ Dalitz plot.

The results are summarized in \tabref{summary}.
All measured \CP\ asymmetries are consistent with zero.
The branching fraction result for \BptochiczKp\ is consistent with the world
average,
while that for \BptoKstarIppiz\ is consistent with and more precise than our
previous measurement~\cite{Aubert:2005cp}, which it supersedes.

\renewcommand{\arraystretch}{1.5}

\begin{table*}[htb]
  \caption{
    Summary of measurements of branching fractions (averaged over charge
    conjugate states) and \CP\ asymmetries.  Both product branching
    fractions and those corrected for secondary decays are shown.  For each
    result, the first uncertainty is statistical, the second is
    systematic and the third, where quoted, is the error on
    $\chicz\to\piz\piz$. The notation $Rh$ refers, where applicable, to the
    intermediate state of a resonance and a bachelor hadron. }
    \label{tab:summary}
\begin{tabular}{lccc}
\hline
Mode            & $\BR(\Bp \to Rh \to \Kppizpiz)$ & $\BR(\Bp \to Rh)$ & $A_{\CP}$ \\
\hline
\BptoKppizpiz   & \kpipiBFal                      & $\cdots$          & \kpipiAcp \\
\hline
\BptoKstarIppiz & \kstarProdBF                    & \kstarBF          & \kstarAcp \\
\BptofIKp       & \fzeroProdBF                    & $\cdots$          & \phantom{$-$}\fzeroAcp \\
\BptochiczKp    & \chiczProdBF                    & \chiczBF          & \chiczAcp \\
\hline
\end{tabular}
\end{table*}

\renewcommand{\arraystretch}{1}

We are grateful for the 
extraordinary contributions of our \pep2\ colleagues in
achieving the excellent luminosity and machine conditions
that have made this work possible.
The success of this project also relies critically on the 
expertise and dedication of the computing organizations that 
support \babar.
The collaborating institutions wish to thank 
SLAC for its support and the kind hospitality extended to them. 
This work is supported by the
US Department of Energy
and National Science Foundation, the
Natural Sciences and Engineering Research Council (Canada),
the Commissariat \`a l'Energie Atomique and
Institut National de Physique Nucl\'eaire et de Physique des Particules
(France), the
Bundesministerium f\"ur Bildung und Forschung and
Deutsche Forschungsgemeinschaft
(Germany), the
Istituto Nazionale di Fisica Nucleare (Italy),
the Foundation for Fundamental Research on Matter (The Netherlands),
the Research Council of Norway, the
Ministry of Education and Science of the Russian Federation, 
Ministerio de Ciencia e Innovaci\'on (Spain), and the
Science and Technology Facilities Council (United Kingdom).
Individuals have received support from 
the Marie-Curie IEF program (European Union), the A. P. Sloan Foundation (USA) 
and the Binational Science Foundation (USA-Israel).

\bibliography{references}

\begin{thebibliography}{33}
\bibitem[{cc()}]{cc}
\bibinfo{note}{{The inclusion of charge conjugate modes is implied throughout
  this paper.}}
\end{thebibliography}
\bibliographystyle{apsrev}

\end{document}